\documentclass[10pt,conference,twocolumn]{IEEEtran}
\usepackage{ifthen}
\usepackage[cmex10]{amsmath}
\usepackage{srcltx}
\usepackage{graphicx}
\usepackage{amssymb}
\usepackage{url}
\usepackage{amsmath}   

\usepackage{array}
\usepackage{epsfig}
\usepackage{latexsym}

\begin{document}


\title{Efficient LDPC Codes over GF($q$) for Lossy Data Compression}


\author{\authorblockN{Alfredo Braunstein}
        \authorblockA{Politecnico di Torino\\
            Dipartimento di Fisica\\
            Corso Duca Degli Abruzzi 24,\\
            10129, Torino, Italy\\
            Email:alfredo.braunstein@polito.it}\and
        \authorblockN{Farbod Kayhan}
        \authorblockA{Politecnico di Torino \\
              Dipartimento di Elettronica\\
              Corso Duca Degli Abruzzi 24,\\
              10129, Torino, Italy\\
              Email: farbod.kayhan@polito.it}\and
        \authorblockN{Riccardo Zecchina}
        \authorblockA{Politecnico di Torino\\
            Dipartimento di Fisica\\
            Corso Duca Degli Abruzzi 24,\\
            10129, Torino, Italy\\
            Email:riccardo.zecchina@polito.it}}

\newtheorem{Theorem}{Theorem}[section]
\newtheorem{Lemma}{Lemma}[section]
\newtheorem{Definition}{Definition}[section]
\newtheorem{Conjecture}{Conjecture}[section]
\newtheorem{Corollary}{Corollary}[section]
\newtheorem{Proposition}{Proposition}[section]

\newcommand{\C}{\mathcal{C}}
\newcommand{\bfx}{\mathbf{x}}
\newcommand{\bfc}{\mathbf{c}}
\newcommand{\bfu}{\mathbf{u}}
\newcommand{\bfs}{\mathbf{s}}
\newcommand{\bfy}{\mathbf{y}}
\newcommand{\bfmu}{\boldsymbol{\mu}}

\newcommand{\bin}[2]{
    \left (
        \begin{array}{@{}c@{}}
        #1 \\ #2
        \end{array}
    \right )
}

\maketitle

\begin{abstract}

In this paper we consider the lossy compression of a binary
symmetric source. We present a scheme that provides a low complexity
lossy compressor with near optimal empirical performance. The
proposed scheme is based on $b$-reduced ultra-sparse LDPC codes
over GF($q$). Encoding is performed by the \emph{Reinforced Belief
Propagation} algorithm, a variant of Belief Propagation. The computational
complexity at the encoder is $\mathcal{O}(<d>.n.q.\log_{2} q)$, where $<d>$ is
the average degree of the check nodes. For our code ensemble, decoding
can be performed iteratively following the inverse steps of the leaf removal
algorithm. For a sparse parity-check matrix the number of needed operations
is $\mathcal{O}(n)$.

\end{abstract}

\IEEEpeerreviewmaketitle

\section{Introduction}
\label{sec:INTR} In this paper we address lossy compression of a
binary symmetric source. Given any realization $\bfy \in \{ 0,1 \}^n $
of a Ber($\frac{1}{2}$) source $\mathbf{Y}$, the goal is to compress
$\bfy$ by mapping it to a shorter binary vector such that an
approximate reconstruction of $\bfy$ is possible within a given
fidelity criterion. More precisely, suppose $\bfy$ is mapped to the
binary vector $\bfx \in \{ 0,1 \}^k$ with $k<n$ and $\hat{\bfy}$ is
the reconstructed source sequence. The quantity $R =\frac{k}{n}$ is
called the compression rate. The fidelity or distortion is measured by
the Hamming distance $d_H (\bfy,\hat{\bfy}) = \frac{1}{n}
\sum_{i=1}^{n} | y_i - \hat{y}_i|$. The goal is to minimize the
average Hamming distortion $D = \mathbb{E} [d_H
  (\mathbf{Y},\hat{\mathbf{Y}})]$ for any given rate. The asymptotic
limit, known as the rate-distortion function, is given by $R(D) = 1 -
H(D)$ for any $D \in [0,0.5]$ where $H(D) = -D\log_2 D - (1-D) \log_2
(1-D)$ is the binary entropy function.

Our approach in this paper is based on Low-Density Parity-Check (LDPC)
codes. Let $\mathcal{C}$ be a LDPC code with $k \times n$ generator
matrix $\mathbf{G}$ and $m \times n$ parity check matrix $\mathbf{H}$.
Encoding in lossy compression can be implemented like decoding in
error correction. Given a source sequence $\bfy$, we look for a
codeword $\hat{\bfy} \in \C$ such that $d_H (\bfy,\hat{\bfy})$ is
minimized. The compressed sequence $\bfx$ is obtained as the $k$
information bits that satisfies $\hat{\bfy} = \mathbf{G}^T \bfx$.

Even though LDPC codes have been successfully used for various types
of lossless data compression schemes \cite{CSV}, and also the
existence of asymptotically capacity-achieving ensembles for binary
symmetric sources has been proved \cite{LDPC-Quantizer}, they have not
been fully explored for lossy data compression. It is partially due to
the long standing problem of finding a practical source-coding
algorithm for LDPC codes, and partially because Low-Density Generator
Matrix (LDGM) codes, as dual of LDPC codes, seemed to be more adapted
for source coding and received more attention in the few past years.

In \cite{MartinianYedidia}, Martinian and Yedidia show that quantizing
a ternary memoryless source coding with erasures is dual of the
transmission problem over a binary erasure channel.  They also prove
that LDGM codes, as dual of LDPC codes, combined with a modified
Belief Propagation (BP) algorithm can saturate the corresponding
rate-distortion bound. Following their pioneering work, LDGM codes
have been extensively studied for lossy compression by several
researchers \cite{lowerboundLDGM,F&F,LDGM-Urbank-LB,LDPC-LDGM,M-W2},
\cite{WainWright}. In a series of parallel work, several researches
have used techniques from statistical physics to provide non-rigorous
analysis of LDGM codes \cite{Ciliberti}, \cite{Kaba1} and
\cite{Murayama}.

In terms of practical algorithms, lossy compression is still an active
research topic. In particular, an asymptotically optimal low
complexity compressor with near optimal empirical performance has not
been found yet.  Almost all suggested algorithms have been based on
some kind of decimation of BP or SP which suffers a computational
complexity of $\mathcal{O}(n^2)$ \cite{Ciliberti}, \cite{F&F} and
\cite{WainWright}. One exception is the algorithm proposed by Murayama
\cite{Murayama}. When the generator matrix is ultra sparse, the
algorithm was empirically shown to perform very near to the associated
capacity needing $\mathcal{O}(n)$ computations. A generalized form of
this algorithm, called reinforced belief propagation (RBP)
\cite{RBP-Blackwell}, was used in a dual setting, for ultra sparse
LDPC codes over GF($2$) for lossy compression \cite{KayhanDP}. The
main drawback in both cases is the non-optimality of ultra sparse
structures over GF($2$) \cite{lowerboundLDGM}, \cite{LDGM-Urbank-LB},
\cite{Murayama}. As we will see, this problem can be overcome by
increasing the size of the finite field.

Our simulation show that $b$-\emph{reduced} ultra sparse LDPC codes
over GF($q$) achieve near capacity performance for $q \geq
64$. Moreover, we propose an efficient encoding/decoding scheme
based on RBP algorithm.

The rest of this paper is organized as follows. Section
\ref{sec:LDPCGF(q)} reviews the code ensemble which we use for lossy
compression. Section \ref{sec:RBPGF($q$)} describes the RBP algorithm
over GF($q$). We also discuss briefly the complexity and
implementation of the RBP algorithm. In section \ref{sec:ILC} we
describe iterative encoding and decoding for our ensemble and then
present the corresponding simulation results in section
\ref{sec:RESULT}. A brief discussion on further research is given in
Section \ref{sec:FR}.

\section{LDPC codes over GF($q$)}
\label{sec:LDPCGF(q)}


In this section we introduce the ultra sparse LDPC codes over
GF($q$). As we will see later, near capacity lossy compression is
possible using these codes and BP-like iterative algorithms.

\subsection{($\lambda,\rho$) Ensemble of GF($q$) LDPC codes}
We follow the methods and notations in \cite{irrLDPC} to construct
irregular bipartite factor graphs. What distinguishes GF($q$) LDPC
codes from their binary counterparts is that each edge ($i,j$) of
the factor graph has a label $h_{i,j} \in $ GF($q$) $ \setminus \{ 0
\}$. In other words, the non-zero elements of the parity-check
matrix of a GF($q$) LDPC codes are chosen from the non-zero elements
of the field GF($q$). Denoting the set of variable nodes adjacent to
a check node $j$ by $\mathcal{N}(j)$, a word $\bfc$ with components
in GF($q$) is a codeword if at each check node $j$ the equation
$\sum_{i \in \mathcal{N}(j)} h_{i,j} c_i = 0$ holds.


A ($\lambda,\rho$) GF($q$) LDPC code can be constructed from a
($\lambda,\rho$) LDPC code by random independent and identically
distributed selection of the labels with uniform probability from
GF($q$)$\setminus \{ 0 \}$ (for more details see \cite{NBLDPC1}).


\subsection{Code Construction for Lossy Compression}
It is well known that the parity check matrix of a GF($q$) LDPC code,
optimized for binary input channels, is much sparser than the one of a
binary LDPC code with same parameters \cite{NBLDPC1,Davey-MacKay}.  In
particular, when $q \geq 2^6$, the best error rate results on binary
input channels is obtained with the lowest possible variable node
degrees, i.e., when almost all variable nodes have degree two.  Such
codes have been called \emph{ultra sparse} or \emph{cyclic} LDPC codes
in the literature. In the rest of this paper we call a LDPC code ultra
sparse (US) if all variable nodes have degree two and the parity
check's degree distribution is concentrated for any given rate. It is
straightforward to show that for a US-LDPC code defined as above check
node degrees has at most two non-zero values and the maximum check
node degree of the code is minimized.

Given a linear code $\C$ and an integer $b$, a $b$-reduction of $\C$
is the code obtained by randomly eliminating $b$ parity-check nodes of
$\C$. For reasons to be cleared in section \ref{sec:ILC}, we are
mainly interested in $b$-reduction of GF($q$) US-LDPC codes for small
values of $b$ ($1 \leq b \leq 5$). Note that by cutting out a parity
check node from a code, the number of codewords is doubled. This
increment of the codewords has an asymptotically negligible effect on
the compression rate since it only increases by $1/n$ while the
robustness may increase.

GF($q$) US-LDPC codes have been extensively studied for transmission
over noisy channels \cite{HuElef}, \cite{DecodingNBLDPC},
\cite{Davey-MacKay}. The advantage of using such codes is twofold. On
the one hand, by moving to sufficiently large fields, it is possible
to improve the code.

On the other hand, the extreme sparseness of the factor graph is
well-suited for iterative message-passing decoding algorithms. Despite
the state of the art performance of moderate length GF($q$) US-LDPC
channel codes, they have been less studied for lossy compression. The
main reason being the lack of fast suboptimal algorithms. In the next
section we present RBP algorithm over GF($q$) and then show that practical
encoding for lossy compression is possible by using RBP as the encoding algorithm for
the ensemble of $b$-reduced US-LDPC codes.

\section{Reinforced Belief Propagation Algorithm in GF($q$)}
\label{sec:RBPGF($q$)} In this section first we briefly review the
RBP equations over GF($q$) and then we discuss in some details the
complexity of the algorithm following Declercq and Fossorier
\cite{DecodingNBLDPC}.

\subsection{BP and RBP Equations}

The GF($q$) Belief Propagation (BP) algorithm is a straightforward
generalization of the binary case, where the messages are
q-dimensional vectors.

Let $\bfmu_{vf}^{\ell}$ denotes the message vector form variable node
$v$ to check node $f$ at the $\ell$th iteration. For each symbol $a
\in $GF($q$), the $a$th component of $\bfmu_{vf}^{\ell}$ is the
probability that variable $v$ takes the value $a$ and is denoted by
$\bfmu_{vf}^{\ell}(a)$. Similarly, $\bfmu_{fv}^{\ell}$ denotes the
message vector from check node $f$ to variable node $v$ at the
iteration $\ell$ and $\bfmu_{fv}^{\ell}(a)$ is its $a$th
component. Also let $\mathcal{N}(v)$ ($\mathcal{M}(f)$) denote the set
of check (variable) nodes adjacent to $v$ ($f$) in a given factor
graph.

Constants $\bfmu_{v}^{1}$ are initialized according to the prior information.
The BP updating rules can be expressed as follows:

{\bf Local Function to Variable:}
\begin{equation}\label{SPfunc2var}
    \bfmu_{fv}^{\ell}(a) \propto \sum_{{\textrm{Conf}}_{(v,f)}(a)}
    \;\; \prod_{ v' \in \mathcal{M}(f) \setminus \{v\}} \bfmu_{v'
    f}^{\ell}(a)
\end{equation}

{\bf Variable to Local Function:}
\begin{equation}\label{SPvar2func}
    \bfmu_{vf}^{\ell+1}(a) \propto \bfmu_{v}^{1}(a)
    \prod_{f' \in \mathcal{N}(v) \setminus
    \{f\}} \bfmu_{f'  v}^{\ell}(a)
\end{equation}

where ${\textrm{Conf}}_{(v,f)}(a)$ is the set of all
configurations of variables in $\mathcal{M}(f)$ which satisfy the
check node $f$ when the value of variable $v$ is fixed to $a$.
We define the marginal function of variable $v$ at iteration $\ell+1$
as

\begin{equation}\label{LocalField}
    \mathbf{g}_{v}^{\ell+1}(a) \propto  \bfmu_{v}^{1}(a) \prod_{f \in \mathcal{N}(v)} \bfmu_{fv}^{\ell}(a).
\end{equation}

The algorithm converges after $t$ iterations if and only if for all
variables $v$ and all function nodes $f$
$$ \bfmu_{fv}^{t+1} = \bfmu_{fv}^{t}$$ up to some precision
$\epsilon$. A predefined maximum number of iterations $\ell_{\max}$
and the precision parameter $\epsilon$ are the input to the algorithm.

RBP is a generalization of BP in which the messages from variable
nodes to check nodes are modified as follows
\begin{equation}
\label{Eq:RBP} \bfmu_{vf}^{\ell+1}(a) \propto
\big(\mathbf{g}_{v}^{\ell}(a) \big)^{\gamma(\ell)}
     \bfmu_{v}^{1}(a) \prod_{f' \in \mathcal{N}(v) \setminus
    \{f\}} \bfmu_{f'  v}^{\ell}(a),
\end{equation}
where $\mathbf{g}_{v}^{\ell}$ is the marginal function of variable $v$
at iteration $\ell$ and $\gamma(\ell):[0,1]\longrightarrow [0,1]$ is
a non-decreasing function. 
Also the equation for each marginal function is changed as below 
\begin{equation}\label{Eq:LocalField-RBP}
    \mathbf{g}_{v}^{\ell+1}(a) \propto \big( \mathbf{g}_{v}^{\ell}(a) \big) ^{\gamma(\ell)} \bfmu_{v}^{1}(a) \prod_{f \in \mathcal{N}(v)} \bfmu_{fv}^{\ell}(a).
\end{equation}

It is convenient to define $\gamma$ to be
\begin{equation*}
\label{coolingfunction} \gamma(\ell) = 1 -
\gamma_{0}\gamma_{1}^{\ell},
\end{equation*}
where $\gamma_{0},\gamma_{1}$ are in $[0,1]$. Note that when
$\gamma_{1} = 1$, RBP is the same as the algorithm presented in
\cite{Murayama} for lossy data compression. In this case it is easy to
show that the only fixed points of RBP are configurations that satisfy
all the constraints.


\subsection{Efficient Implementation}

Ignoring the normalization factor in (\ref{SPvar2func}), to compute
all variable to check-node messages at a variable node of degree $d_v$
we need $\mathcal{O} (q.d_{v})$ computations. A naive
implementation of GF($q$) BP has computational complexity of
$\mathcal{O}(d^{2}_{f}.q^2)$ operations at each check node of degree
$d_{f}$. This high complexity is mainly due to the sum in
(\ref{SPfunc2var}), that can be interpreted as a discrete convolution
of probability density functions. Efficient implementations of
function to variable node messages based on Discrete Fourier Transform
have been proposed by several authors, see for example
\cite{R&U,MacKay,NBLDPC1}, \cite{DecodingNBLDPC} and the references
within. The procedure consists in using the identity $\bigodot_{v'
  \in \mathcal{M}(f) \setminus \{v\}}\bfmu_{v'f} =
\mathcal{F}^{-1}\left(\prod_{ v' \in \mathcal{M}(f) \setminus \{v\}}
\mathcal{F}\left(\bfmu_{v'f}\right)\right)$, where the symbol
$\bigodot$ denotes convolution.

Assuming $q=b^p$, the Fourier transform of each message $\bfmu_{v'f}$
needs $\mathcal{O}(q.p)$ computations and hence the total
computational complexity at check node $f$ can be reduced into
$\mathcal{O}(d^{2}_{f}.q.p)$. This number can be further reduced to
$\mathcal{O}(d_{f}.q.p)$ by using the fact that $\prod_{ v' \in
  \mathcal{M}(f) \setminus \{v\}} \mathcal{F}\left(\bfmu_{vf}\right) =
\prod_{ v' \in \mathcal{M}(f)}
\mathcal{F}\left(\bfmu_{v'f}\right)/\mathcal{F}\left(\bfmu_{vf}\right)$,
or alternatively by using the summation strategy described in
\cite{BrMuPa} which has the same complexity but is numerically more
stable. Therefore, the total number of computations per iteration is
$\mathcal{O}(<d>.q.p.n)$ where $<d>$ is the average degree.
\section{Iterative Lossy Compression}
\label{sec:ILC}

In the following three subsections we first describe a simple method
for identifying information bits of a $b$-reduced US-LDPC code and
then present a near capacity scheme for iterative compression
(encoding) and linear decompression (decoding).

\subsection{Identifying a Set of Information Bits}

For $b$-reduced US-LDPC codes, one can use the \emph{leaf removal} (LR)
algorithm to find the information bits in a linear time. In the rest
of this section we briefly review the LR algorithm and show that
1-reduction (removal of a sole check node) of a US-LDPC code
significantly changes the intrinsic structure of the factor graph of
the original code.

The main idea behind LR algorithm is that a variable on a leaf of a
factor graph can be fixed in such a way that the check node to which
it is connected is satisfied \cite{CoreRZ}. Given a factor graph, LR
starts from a leaf and removes it as well as the check node it is
connected to. LR continues this process until no leaf remains. The
residual sub-graph is called the \emph{core}.  Note that the core is
independent of the order in which leaves (and hence the corresponding
check nodes) are removed from the factor graph. This implies that also
the number of steps needed to find the core does not depend on the
order on which leaves are chosen.

While US-LDPC codes have a complete core, i.e. there is no leaf in
their factor graph, the $b$-reduction of these codes have empty core.
Our simulations also indicate that even 1-reduction of a code largely
improves the encoding under RBP algorithm (see section
\ref{sec:RESULT}).  How RBP exploits this property is the subject of
ongoing research. It is straightforward to show that a code has empty core
if and only if there exists a permutation of columns of the corresponding 
parity-check matrix $\mathbf{H}$ such that $h_{ij} \neq 0$ for $i=j$ and 
$h_{ij} = 0$ for all $i > j$.  

As we have mentioned, LR algorithm can be also used to find a set of
information bits of a given US-LDPC code. At any step $t$ of LR
algorithm, if the chosen leaf is the only leaf of the check node $f_t$
into which it is connected, then its value is determined uniquely as a
function of non-leaf variables of check node $f_t$. If the number of
leaves $d_t$ is greater than 1, there are $2^{d_t-1}$ configurations
which satisfy the check node after fixing the values of non-leaf
variables.  At each step of LR we choose a subset of $d_t-1$ leaves.
This set is denoted by $F_{t}^{LR}$ and we call it the free subset at
$t^{th}$ step. Note that there are $d_t$ free subsets among which we
choose only one at each step.  It is straightforward to show that the
union of all free subsets $F=\cup_t F_t^{LR}$ is a set of information
bits for a given US-LDPC code.

\subsection{Iterative Encoding} \label{subsec:Encoding} Suppose a code
of rate $R$ and a source sequence $\bfy$ is given. In order to find
the codeword $\hat{\bfy}$ that minimizes $d_H(\hat{\bfy},\bfy)$, we
will employ the RBP algorithm with a strong prior
$\bfmu^1_{v}(a)=\exp(-L d_H(y_v,a))$ centered around $\bfy$. The sequence
of information bits of $\hat{\bfy}$ is the compressed sequence and is denoted by $\bfx$.
In order to process the encoding in GF($q$), we
first need to map $\bfy$ into a sequence in GF($q$). This can be
simply done by grouping $b$ bits together and use the binary
representation of the symbols in GF($q$).

\subsection{Linear Decoding} \label{subsec:Decoding}

Given the sequence of information bits  $\bfx$, the goal of the
decoder is to find the corresponding codeword $\hat{\bfy}$.  This can
be done by calculating the $\mathbf{G}^T \bfx$ which in general needs
$\mathcal{O}(n^2)$ computations. One of the advantages of our scheme
is that it allows for a low complexity iterative decoding. The
decoding can be performed by iteratively fixing variables following
the inverse steps of the LR algorithm; at each step $t$
only one non-information bit is unknown and its value can be
determined from the parity check $f_t$.  For a sparse parity-check
matrix, the number of needed operations is $\mathcal{O}(n)$.


\section{Simulation Results} \label{sec:RESULT}
\subsection{Approximating the Weight Enumeration Function by BP}

Given an initial vector $\bfy$, and a probability distribution $P(\bfc)$
over all configurations, the $P$-average distance from $\bfy$ can be
computed by

\begin{equation}
D_{P}(\bfy) = \sum_{i}\sum_{c_i} P(c_i) d_{H}(c_i,y_i)\label{eq:avd}
\end{equation}

where $P(c_i)$ is the set of marginals of $P$. On the other hand, the
entropy of the distribution $P$ is defined by

\begin{equation}
S(P) = - \sum_{\bfc} P(\bfc) \log P(\bfc)\label{eq:S}.
\end{equation}

Even though it is a hard problem to calculate analytically both
marginals and $S(P)$ of a given code, one may approximate them using
messages of the BP algorithm at a fixed point \cite{Yedida}. Assuming
the normalized distance is asymptotically a self-averaging quantity
for our ensemble, $S(P)$ represents the logarithm of the number of
codeword at distance $D_{P}(\bfy)+\mathcal{O}(1)$ from $\bfy$. By applying a prior
distribution on codewords given by $\exp(-L d_H(\bfc,\bfy))$ one is able
to sample the sub-space of codewords at different distances from
$\bfy$.

\begin{figure}[h]
\begin{centering}
    \includegraphics[angle=0,width=0.45\textwidth]{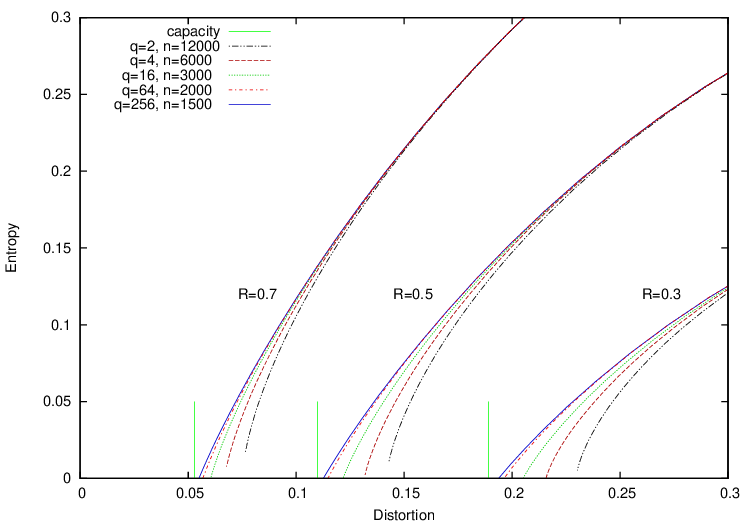}
    \caption {\label{Fig:WEF} The approximate WEF of GF($q$) US-LDPC
    codes as a function of $q$ for a same blocklength in binary digits. }
\end{centering}
\end{figure}

Fig. \ref{Fig:WEF} demonstrates the WEF of random GF(q) US-LDPC codes
for rates 0.3, 0.5, and 0.7 and field orders 2, 4, 16, 64 and 256. The
blocklength is normalized so that it corresponds to $n=12000$ binary
digits.

Though BP is not exact over loopy graphs, we conjecture that the WEF
calculated for US-LDPC codes is asymptotically exact. This hypothesis
can be corroborated by comparing the plot in Fig. \ref{Fig:WEF} with
the simulation results we obtained by using RBP algorithm (Fig.
\ref{Fig:qPerformace}).

\subsection{Performance}
In all our simulations the parameter $\gamma_1$ of RBP algorithm is
fixed to one and therefore the function $\gamma$ is constant and
does not depend on the iterations. We also fix the maximum number of
iterations into $\ell_{max} =300$. If RBP does not converge after
300 iterations, we simply restart RBP with a new random scheduling.
The maximum number of trials allowed in our simulations is $T_{max}
=5$. The encoding performance depends on several parameters such as
$\gamma_0$,  $L$, the field order $q$, and the
blocklength $n$. In the following we first fix $n$, $q$ and
$L$, in order to see how the performance changes as a
function of $\gamma_0$.
\subsubsection{Performance as a Function of $\gamma_0$}
\label{subsubsec:performancePEG} Our main goal is to show that there
is a trade off, controlled by $\gamma_0$, between three main aspects
of the performance, namely: average distortion, average number of
iterations and average number of trials. The simulations in this
subsection are done for a 5-reduced GF(64) US-LDPC code with length
$n=1600$ and rate $R=0.33$. The factor graph is made by
\emph{Progressive-Edge-Growth} (PEG) construction \cite{HuElef}. The
rate is chosen purposefully from a region where our scheme has the
weakest performance. The distortion capacity for this rate is
approximately $0.1754$.

\begin{figure}[h]
\begin{centering}
    \includegraphics[angle=0,width=0.45\textwidth]{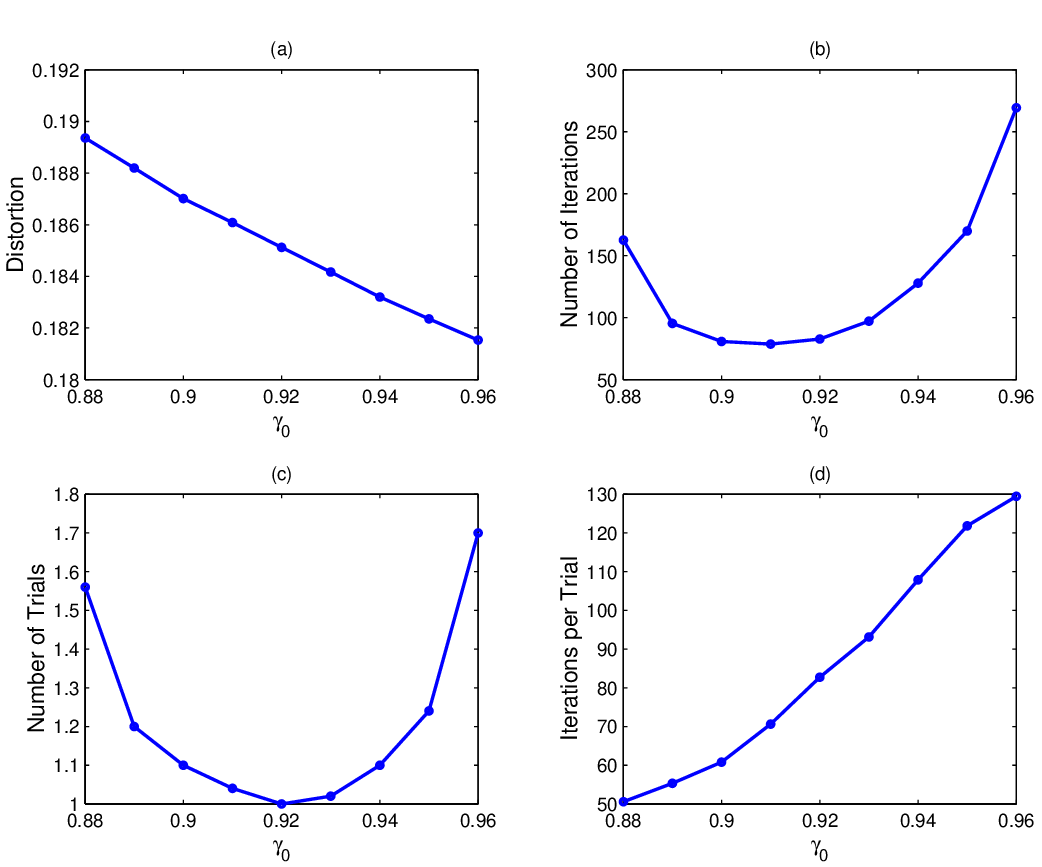}
    \caption {\label{Fig:gamma0} Performance as a function of $\gamma_0$
    for a PEG graph with n=1600 and R=0.33. The averages are taken
    over 50 samples.(a) Average distortion as a function of $\gamma_0$.
    For $\gamma_0 > 0.96$ the RBP
    does not converge within 300 iterations. (b)The average number of
    iterations. (c)The average number of trials.  (d) The average number of iterations
    needed for each trial. Note that even though average number of iterations
    show a steep increase as a function of $\gamma_0$, the average number of
    iterations needed per trial increases only linearly.  }
\end{centering}
\end{figure}

In Fig. \ref{Fig:gamma0} we plot the performance as a function
of $\gamma_0$. For $\gamma_0 = 0.92$ we achieve a distortion of
$D=0.1851$ needing only 83 iterations in average and without any
need to restart RBP for 50 samples. By increasing $\gamma_0$ to 0.96,
one can achieve an average distortion of $0.1815$ which is only 0.15
dB away from the capacity needing 270 iterations in average.

\subsubsection{Performance as a function of $R$ and $q$}

\begin{figure}[h]
\begin{centering}
    \includegraphics[angle=0,width=0.43\textwidth]{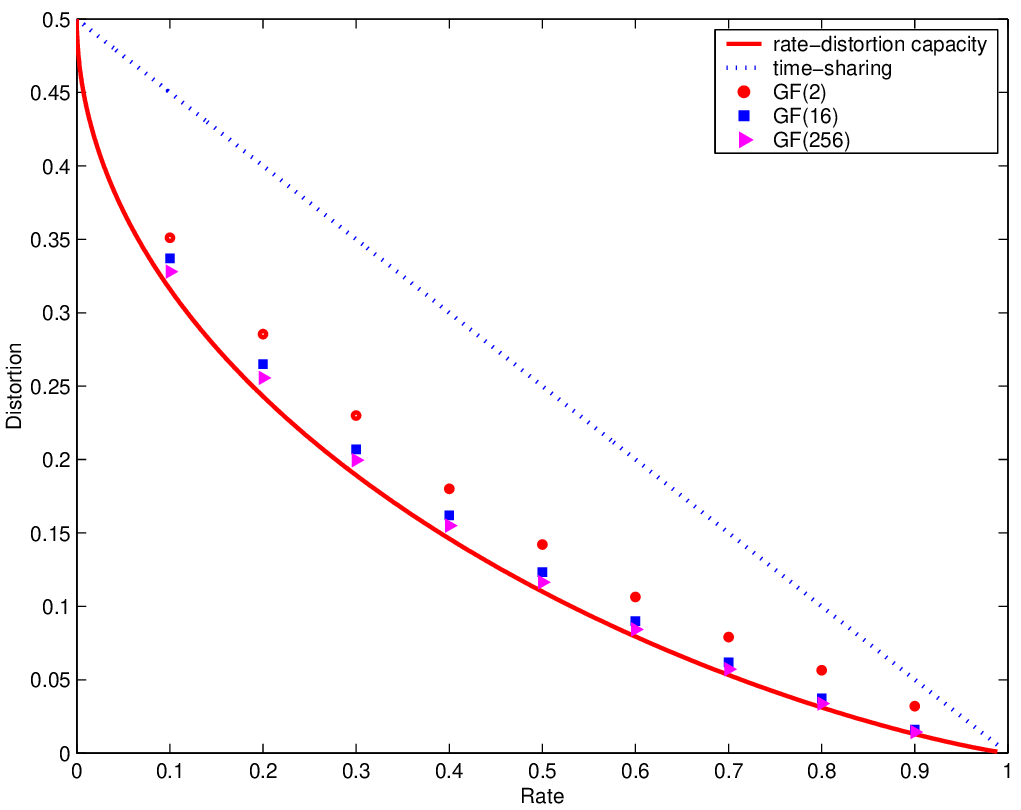}
    \caption {\label{Fig:qPerformace} The rate-distortion performance
    of GF($q$) LDPC codes encoded with RBP algorithm for $q=2, 16$ and $256$.
    The blocklength is
    12000 binary digits and each point is the average distortion over
    50 samples.}
\end{centering}
\end{figure}

Fig. \ref{Fig:qPerformace} shows the distortion obtained by randomly
generated 5-reduced GF(q) US-LDPC codes for $q=2$, $q=16$ and
$q=256$. The block length is fixed to $n=12000$ binary digits. For each
given code, we choose $\gamma_0$ and $L$ so that the average number of
trials does not exceed 2 and the average number of iterations remains
less than 300. Such values of $\gamma_0$ and $L$ are found by
simulations. Under these two conditions, we report distortion
corresponding to best values of the two parameters averaged over 50
samples.

\section{Discussion and Further Research}
\label{sec:FR} Our results indicate that the scheme proposed in this
paper outperforms the existing methods for lossy compression by
low-density structures in both performance and complexity. The main
open problem is to understand and analyze the behavior of RBP over
$b$-reduced US-LDPC codes.

As we have mentioned, $b$-reduction of a US-LDPC code not only
provides us with simple practical algorithms for finding information
bits and decoding, but also largely improves the convergence of RBP.
It is interesting to study the ultra sparse ensembles where a certain
fraction of variable nodes of degree one is allowed.


\section*{Acknowledgment}

F.K. wish to thank Sergio Benedetto, Guido Montorsi and
Toshiyuki Tanaka for valuable suggestions and useful discussions.


\end{document}